\documentclass{aipproc}
\begin{document}

\title{Parallel Computing on a PC Cluster}
\author{X.Q. Luo$^1$, E.B. Gregory$^1$, J. C. Yang$^2$, Y. L. Wang$^2$, D. Chang$^2$, and Y. Lin$^2$}
\affiliation{$^1$Department of Physics, Zhongshan University, 
Guangzhou 510275, China\\
$^2$Guoxun, Ltd, Guangzhou, China}

\begin{abstract}
The tremendous advance in computer technology in the past decade has made 
it possible to achieve the
performance of a supercomputer on a very small budget. We have built a 
multi-CPU cluster of Pentium PC
capable of parallel computations using the Message Passing Interface 
(MPI). We will discuss the
configuration, performance, and application of the cluster to our work 
in physics.
\end{abstract}

\maketitle

\section{Introduction}
The lattice field theory group at the Zhongshan University has faced the 
familiar pressures of trying to balance the need for increased computational 
power against the constraints of an academic research budget.  The group's
primary research interest is in lattice quantum chromodynamics (QCD), the 
study of quark interactions. This field lends itself well to numerical
simulation, but requires significant computational resources for forefront
research. Traditionally this has been the domain of supercomputers. However,
in recent years, advances in technology and falling hardware prices have 
blurred the distinction between the definition of a supercomputer and personal 
computer.  Desktop machines of today are far more powerful than the 
supercomputers of yesteryear.  

A further development is that the fastest 
computers today are in fact parallel computers, with multiple processors 
working together on a problem.  Parallel computation has its limitations, the 
biggest being that it is applicable only to problems that can be divided into 
concurrent tasks.  Furthermore, since communication between processors is 
usually the slowest part of the computation, the separate tasks should 
ideally be as independent as possible.  Fortunately many computational 
physics problems,  including lattice QCD, fall into the category of 
parallelizable problems.  Indeed, many computational problems in the 
commercial world are suitable for this type of computation as well.  
Applications of parallel computation include graphics and animation, 
telecommunications and internet service, and many other fields heavily reliant 
on computer processing.

It is possible to join multiple cheap, fast PC 
type computers to build a parallel ``supercomputer'' with an arbitrarily high
aggregate speed. A cluster of this type is called a ``Beowulf Cluster'' 
\cite{beowulf} 
and the idea was pioneered by the United States' National Aeronautics and  
Space Administration.

\section{Construction}
\subsection{Hardware}
One big advantage of a PC cluster over other types of supercomputers is the 
low cost and easy availability of the hardware components. All the hardware 
in our cluster is available at retail computer suppliers. This gives great 
flexibility in both building and the cluster and in any future upgrades or 
expansions we may choose to make. 

Our cluster consists of ten PC type computers, each with two 500 MHz 
Pentium III processors inside.  The logic behind dual CPU machines is that one
can double the number of processors without the expense of additional, cases,
power supplies, motherboards, network cards, et cetera. Also, the inter-node 
communication speed is faster for each pair of processors in the same box
as compared to communication between separate computers. Each computer has 
an 8GB EIDE hard drive, 128 MB of memory, a 100Mbit/s ethernet card, a simple
graphics card a floppy drive and a CDROM.  In practice the CDROM, the floppy 
drive, and even the graphics card could be 
considered extraneous,  as all interactions with the nodes could be done 
through the network.  However, with these 
components, all of which are relatively cheap in comparison to the total cost, 
the operating system installation and occasional maintenance is 
significantly easier.
One computer has a larger hard disk (20 GB), and a SCSI card for 
interaction with a tape drive.  For the entire cluster we have only one console
consisting of a keyboard, mouse and monitor.

\begin{figure}
\resizebox{\columnwidth+\columnwidth}{!}{\rotatebox{-90}{\includegraphics{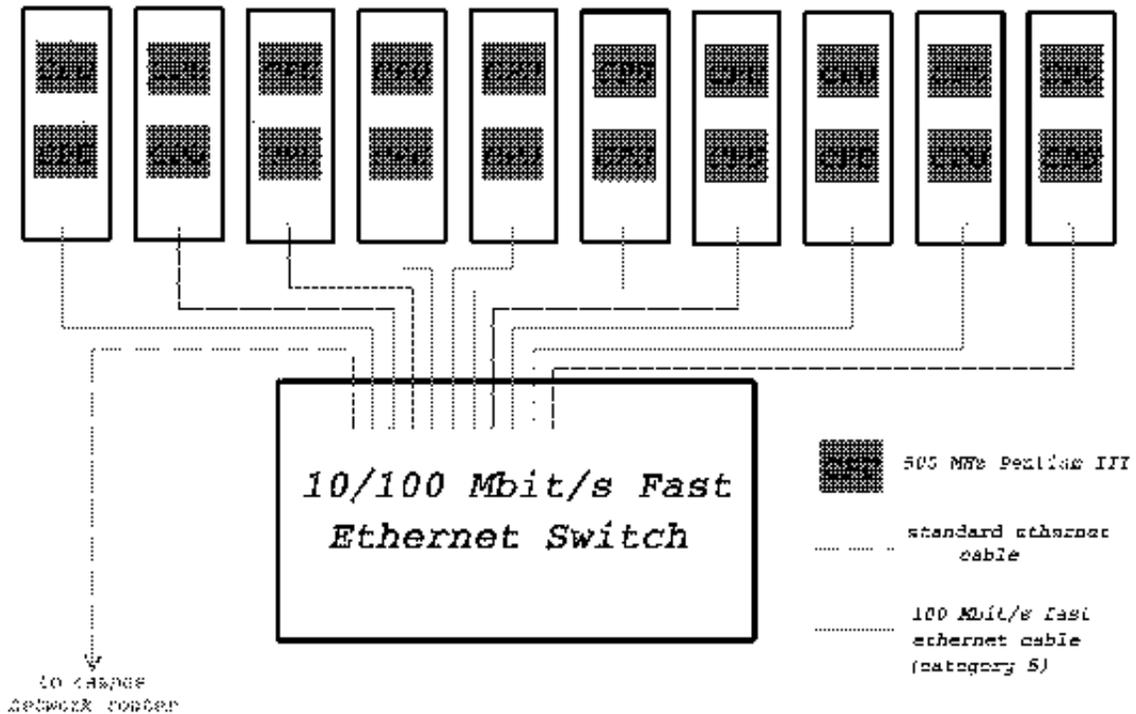}}}
\caption{The layout of a 10 dual-CPU node cluster.}
\label{fig:a}
\end{figure}

A fast ethernet switch handles the inter-node communication. The switch has 
24 ports so there is ample room for future expansion of the cluster to up to 
a total of 48 processors. Of course it is possible to link multiple switches
or use nodes with more that two processors, so the possibilities for a larger 
cluster are nearly limitless. The layout of the cluster is illustrated in 
Figure \ref{fig:a}.
 
\subsection{Software}
The cluster runs on the Linux operating system.  Linux is powerful and 
inexpensive. It easily supports important features like multiple processors. 
It allows the configuration of a network file server. We have mounted 
the largest hard disk on to all of the machines in the cluster. Each machine
can read 
and write to it as if it were physically part of that computer. Linux also 
supports a network information system to share user accounts across the 
entire cluster.  One uses the same account and home directory, no matter 
which machine he or she logs into. 
Standard Linux distributions also supply C, C++, and Fortran compilers.

We can use the the cluster for parallel processing by using the message 
passing interface (MPI)\cite{mpi}, 
a library of communication functions and programs 
that allow for communication between processes on different CPUs.
The programmer must design the parallel algorithm so that it appropriately
divides the task among the individual 
processors. He or she must then include message passing 
functions in the code which allow information to be sent and received by the 
various processors.  MPI is one of the most popular standards for message 
passing parallel programming, and is widely used in the physics community.
Therefore we are able to share parallel programs in C, C++, or Fortran with 
collaborators elsewhere in the world who may even be running MPI on a 
different  platform.

\section{Performance}
\subsection{Serial Benchmark}
We have run the LINPACK benchmark \cite{linpack}, a standard serial benchmark 
test on our computers to measure the speed of a single processor. 
The benchmark showed that a single 500 MHz Pentium III processor is capable 
of a peak speed between 84 Mflops and 114 Mflops (million floating point 
operations per second) for single precision arithmetic and between 62 Mflop 
and  68 Mflop for double precision arithmetic.  The peak aggregate speed for
the entire cluster of twenty processors, is therefore about 2 Gflops.

\begin{table}
\begin{tabular}{ccc} 
\hline 
\raisebox{0pt}[12pt][6pt]{Machine} & 
\raisebox{0pt}[12pt][6pt]{$\mu$-sec/link} & 
\raisebox{0pt}[12pt][6pt]{MB/sec}\\
\hline
\raisebox{0pt}[12pt][6pt]{SX-4} & 
\raisebox{0pt}[12pt][6pt]{4.50} & 
\raisebox{0pt}[12pt][6pt]{45}\\

\raisebox{0pt}[12pt][6pt]{SR2201} & 
\raisebox{0pt}[12pt][6pt]{31.4} & 
\raisebox{0pt}[12pt][6pt]{28}\\
\raisebox{0pt}[12pt][6pt]{Cenju-3} & 
\raisebox{0pt}[12pt][6pt]{57.42} & 
\raisebox{0pt}[12pt][6pt]{8.1}\\
\raisebox{0pt}[12pt][6pt]{Paragon} & 
\raisebox{0pt}[12pt][6pt]{149} & 
\raisebox{0pt}[12pt][6pt]{9.0}\\
\raisebox{0pt}[12pt][6pt]{\bf ZSU's Pentium cluster} & 
\raisebox{0pt}[12pt][6pt]{\bf 7.3} & 
\raisebox{0pt}[12pt][6pt]{\bf 11.5}\\
\hline

\end{tabular}
\caption{Comparison of performance of MPI QCD benchmark. Comparison data
from Hioki and Nakamura. \cite{hioki}}\label{table1}
\label{tab:qcdim}
\end{table}

\subsection{QCD Benchmark}
As we primarily developed the cluster for numerical simulations of 
lattice QCD, we have also performed a benchmark which specifically tests the 
performance in a parallel lattice QCD code.  Lattice QCD simulations are well 
suited
for parallelization \cite{gupta} as they involve mostly local calculations 
on a multi-dimensional lattice. 
The algorithm can conveniently divide the lattice and assign the sections
to different processors.  The communication between the nodes therefore 
is not extremely large.
Hioki and Nakamura \cite{hioki}
provide comparison performance data on SX-4 (NEC), SR2201 (Hitachi),
 Cenju-3 (NEC) and  Paragon (Intel) machines.
Specifically, we compare the computing time per link update in microseconds 
per link and the inter-node communication speed in MB/sec. The link update is a
fundamental computational task within the QCD simulation and is therefore a 
useful standard.  The test was a simulation of improved pure gauge lattice
action ($1\times 1$ plaquette and $1\times 2$ rectangle terms) on a 
$16^4$ lattice. In each case the simulation was run on 16 processors.  The 
results are summarised in Table \ref{tab:qcdim}.

\subsection{Cost Comparison}
We believe that such a parallel cluster of PCs  may be the cheapest solution 
to the problem of developing computing resources for scientific simulations.
In 1999, our cluster cost about US\$14,000, including all hardware and 
software.
This equates to roughly \$7/Mflop. We can compare this to a commercial 
supercomputer. The Cray T3E-1200E uses 1.2 Gflop processors \cite{cray}.  The
basic starting model comes with six processors for a total peak speed of 
7.2 Gflops. The cost for the six node model, though,  is US\$630,000, or 
\$87.50/Mflop.  Our home made supercomputer is more than an order of 
magnitude cheaper.  

Of course this is a naive comparison, as the Cray differs in many ways. 
Notably, faster individual processors means serial jobs will run much faster, 
and parallel programs will require fewer processors, and hence less 
inter-processor communication. Furthermore the inter-processor 
communication is much faster on the Cray. 

It is clear, however that for numerical tasks that are easily broken in 
fairly independent tasks, a farm of PCs is an extremely economical solution 
by comparison. Additionally, the PC cluster is highly scalable. PCs and their 
components are so ubiquitous, that expansion of the system is trivial. 
Nearly anyone with a screwdriver, can upgrades or replaceme components so 
it is not nescessary to have a service contract with a commercial vendor.

\section{conclusions}
We feel that our parallel cluster of PC type computers is an example of an 
economical way to build a powerful computing resource for academic purposes.
On an MPI QCD benchmark simulation it compares favorably with other MPI 
platforms. It is also drastically cheaper than commercial supercomputers 
for the same amount of processing speed. PC clusterssuch as this one 
have applications in both academia and in commercial enterprises.
 It is particularly suitable for 
developing research groups in countries where funding for pure research is 
more scarce. We believe that our cluster may be the first such facility at an 
academic physics department in mainland China. 

\section{Acknowledgements}
This work is supported by the
National Science Fund for Distinguished Young Scholars (19825117),
National Science Foundation, Guangdong Provincial Natural Science Foundation (990212) and 
Ministry of Education of China.
We are grateful for generous additional support from Guoxun 
(Guangdong National Communication Network) Ltd.. 
We would 
also like to thank Shinji Hioki of Tezukayama University for the use of the
QCDimMPI code. The C version of the LINPACK benchmark was written by Bonnie 
Toy.

\bibliographystyle{aipproc}

\end{document}